\documentclass[a4paper,11pt]{article}

\usepackage[english]{babel}
\usepackage[a4paper,tmargin=3truecm,bmargin=3truecm,rmargin=2.5truecm,
lmargin=2.5truecm,twoside,verbose=true]{geometry}

\usepackage{cancel,graphicx}
\usepackage[pdftex]{hyperref}

\usepackage{amsmath,amssymb}

\numberwithin{equation}{section}

\allowdisplaybreaks[1]

\renewenvironment{thebibliography}[1]
         {\section*{References}\frenchspacing\small
          \begin{list}{[\arabic{enumi}]}
         {\usecounter{enumi}\parsep=2pt\topsep 0pt
         \settowidth{\labelwidth}{[#1]}
         \leftmargin=\labelwidth\advance\leftmargin\labelsep
         \rightmargin=0pt\itemsep=1pt\sloppy}}{\end{list}}

\title{On the vacuum states for noncommutative gauge theory\footnote{Work 
supported by ANR grant NT05-3-43374 ``GenoPhy''.}}
\author{Axel de Goursac$^{a,b}$, Jean-Christophe Wallet$^a$, 
Raimar Wulkenhaar$^b$}
\date{}

\begin{document}

\maketitle
\vspace*{-1cm}
\begin{center}
\textit{$^a$Laboratoire de Physique Th\'eorique, B\^at.\ 210\\
    Universit\'e Paris XI,  F-91405 Orsay Cedex, France\\
    e-mail: \texttt{axelmg@melix.net}, 
\texttt{jean-christophe.wallet@th.u-psud.fr}}\\[1ex]
\textit{$^b$Mathematisches Institut der Westf\"alischen
  Wilhelms-Universit\"at \\Einsteinstra\ss{}e 62, D-48149 M\"unster,
  Germany \\
    e-mail: \texttt{raimar@math.uni-muenster.de}}\\
\end{center}%

\vskip 2cm

\begin{abstract}
Candidates for renormalisable gauge theory models on Moyal spaces constructed recently have non trivial vacua. We show that these models support vacuum states that are invariant under both global rotations and symplectic isomorphisms which form a global symmetry group for the action. We compute the explicit expression in the position space for these vacuum configurations in two and four dimensions.
\end{abstract}

\pagebreak

\section{Introduction.}

A new family of noncommutative (NC)  field theories \cite{Douglas:2001ba,Szabo,Wulkenhaar:2006si} (For details on Non Commutative Geometry, see \cite{CONNES,CM}) came under increasing scrutiny after 1998 when it was realized \cite{Seiberg:1999vs,Schomerus} that string theory seems to have some effective regimes described by noncommutative field theories (NCFT) defined on a simple NC version of flat four dimensional space. This latter is the so-called Moyal-Weyl space (for a mathematical description see e.g \cite{Gracia-Bondia:1987kw,Varilly:1988jk}) which has constant commutators between space coordinates. NCFT on Moyal spaces were also shown to be the ones whose nonrelativistic counterparts correspond to many body quantum theory in strong external magnetic field (see e.g \cite{hall1,hall2}).

This growing interest received however a blow when it was noticed \cite{Minwalla:1999px,Chepelev:1999tt} that the simplest NC $\varphi^4$ model , ($\varphi$ real-valued) on the 4-dimensional Moyal space is not renormalizable due to the occurence of a phenomenon called Ultraviolet/Infrared mixing  \cite{Minwalla:1999px,Chepelev:1999tt,Matusis:2000jf}. This phenomenon results basically from the existence of some nonplanar diagrams which are ultraviolet finite but nevertheless develop infrared singularities which when inserted into higher order diagrams are not of the renormalizable type  \cite{Douglas:2001ba,Szabo,Wulkenhaar:2006si}. A possible solution to this problem, hereafter called the "harmonic solution", was proposed in 2004 \cite{Grosse:2004yu,Grosse:2003aj} (see also \cite{Rivasseau:2005gd,Gurau:2005gd,Tanasa:2007qx}) which basically amounts to supplement the initial action with a simple harmonic oscillator term leading to a fully renormalisable NCFT. For recent reviews, see e.g \cite{vince,Wallet:2007 em}. This result seems to be related to the covariance of the model under a new symmetry, called Langmann-Szabo duality  \cite{Langmann:2002cc}, which roughly speaking, exchanges coordinates and momenta. Other renormalisable noncommutative matter field theories on Moyal spaces have then been identified  \cite{Grosse:2003nw,Langmann:2003if,Langmann:2003cg,Vignes-Tourneret:2006nb} and some studies of the properties of the corresponding renormalisation group flows have been carried out \cite{beta1,Lakhoua:2007ra,beta23}, exhibiting in particular the vanishing of the $\beta$-function to all orders for the $\varphi^4_4$ model \cite{beta}.

So far, the construction of a fully renormalisable gauge theory on 4-dimensional
Moyal spaces remains a challenging problem. Recall that the naive noncommutative version of the pure Yang-Mills action on Moyal
spaces given by $S_0=\frac 14\int d^4x (F_{\mu\nu}\star F_{\mu\nu})(x)$ (in the standard notations and conventions recalled in section 2), suffers from UV/IR mixing which makes its renormalisability unlikely. This basically stems from the occurence of an IR singularity in the one-loop polarisation tensor $\omega_{\mu\nu}(p)$ ($p$ is some external momentum). Indeed, by standard calculation one easily infers that
\begin{align}
\omega_{\mu\nu}(p)\sim\frac{(D-2)}{4}\Gamma(\frac D2)\frac{\widetilde p_\mu\widetilde p_\nu}{\pi^{D/2}(\widetilde p^2)^{D/2}}+... \quad p\to0 
\end{align}
where $\widetilde p_\mu\equiv 2\Theta^{-1}_{\mu\nu}p_\nu$ and $\Gamma(z)$ denotes the Euler function. Notice that this singularity, albeit obviously transverse in the sense of the Slavnov-Taylor-Ward identities, does not correspond to some gauge invariant term. This implies that the recent alternative solution to the UV/IR mixing proposed within the NC $\varphi^4$ model in \cite{GMRT}, which roughly amounts to balance the IR singularity throught a counterterm having a similar form, cannot be extended straighforwardly (if possible at all) to the case of gauge theories.

It turns out that the extension of the harmonic solution to the case of gauge theories has been achieved recently in \cite{de Goursac:2007gq} and \cite{Grosse:2007qx} (see also  \cite{deGoursac:2007qi,Wohl2007}), starting basically from a computation of the one-loop effective gauge action obtained by integrating out the matter degree of freedom of a NC matter field theory with harmonic term similar to the one used in \cite{Grosse:2004yu,Grosse:2003aj} minimally coupled to an external gauge potential. Both analysis have singled out, as possible candidate for renormalisable gauge theory defined on (4-dimensional) Moyal space, the following generic action
\begin{align}
S=\int d^4x \Big(\frac 14F_{\mu\nu}\star F_{\mu\nu}
+\frac{\Omega^2}{4}\{\mathcal A_\mu,\mathcal A_\nu\}^2_\star
+\kappa\mathcal A_\mu\star\mathcal A_\mu\Big) \label{eq:decadix1}
\end{align}
Here, $\mathcal A_\mu$ denotes the so called covariant coordinates defined in section 2, a natural gauge covariant tensorial form stemming
from the existence of a canonical gauge invariant connection within the present NC framework (for more details on the relevant mathematical structures see e.g in \cite{Dubois-Violette:1989vq, Masson:1999, Wallet:2007 em}. In \eqref{eq:decadix1}, the additional 2nd term may be viewed as a "gauge counterpart'' of the harmonic term introduced in \cite{Grosse:2004yu}. This action, that has been shown to be related to a spectral triple \cite{Grosse:2007jy}, exhibits interesting properties \cite{de Goursac:2007gq,Grosse:2007qx} that deserve further studies. For instance, gauge invariant mass terms for the gauge fields are allowed even in the absence of some Higgs mechanism. However, the presence of the additional terms implies in general a non-vanishing vacuum expectation value for the gauge potential. Somewhat similar non trivial vacuum configurations also occur within NC scalar models with harmonic term as shown and studied recently in \cite{deGoursac:2007uv}. It turns out that the explicit determination of the relevant vacuum for any gauge theory model of the type \eqref{eq:decadix1} is a necessary step to be reached before the study of its renormalisability can be undertaken. Indeed, a reliable pertubative analysis in the present situation can only be defined after the action is expanded around the non trivial vacuum which actually demand to know the explicit expression for that vacuum. This should then be followed by a suitable gauge-fixing presumably inherited from the background field methods.

The purpose of the present paper is the determination of vacuum configurations for the gauge theories generically defined by \eqref{eq:decadix1}, for $D=2$ and $D=4$ dimensions. The paper is organized as follows. In the section 2, we fix the notations and collect the main features of the Noncommutative Geometry framework that will be used thourough the analysis. The relevant symmetries for the vacua are examined in the section 3, focussing on vacuum configurations that are invariant under both rotations and symplectic isomorphisms. The sections 4 and 5 are devoted to the determination of these symmetric vacuum configurations for $D=2$ and $D=4$. It turns out that the use of the matrix base formalism proves convenient to obtain explicit expressions for the relevant solutions of the equation of motion from which vacuum solutions can be obtained. Finally, we summarize and discuss our reults in the section 6 and we conclude.

\section{Basic features.}

In this section, we collect the main ingredients that will be needed in the subsequent discussion. Some detailed studies of the Moyal NC algebra are carried out e.g in \cite{Gracia-Bondia:1987kw,Varilly:1988jk} while mathematical descriptions of the NC framework underlying the present study can be found in \cite{Masson:1999,Wallet:2007 em}.

\subsection{The noncommutative gauge theory.}

The Moyal space, on which is constructed the gauge theory considered here, can be defined as an algebra of tempered distributions on $\mathbb R^D$ endowed with the Moyal product \cite{Gracia-Bondia:1987kw,Varilly:1988jk} hereafter denoted by a $\star$-symbol. Indeed, this latter can be defined on $\mathcal S=\mathcal S(\mathbb R^D)$, the space of complex-valued Schwartz functions by
\begin{align}
\forall f,h\in\mathcal{S}\quad(f\star h)(x)=\frac{1}{\pi^D\theta^D}\int d^Dy\,d^Dz\ 
f(x+y)\,h(x+z)e^{-iy\wedge z},\label{eq:moyal}
\end{align}
where $x\wedge y=2x_\mu\Theta^{-1}_{\mu\nu}y_\nu$ and
\begin{align}
\Theta_{\mu\nu}=\theta\begin{pmatrix} 0 & -1 & 0 & 0 & \\ 1 & 0 & 0 & 0 & \\ 0 & 0 & 0 & -1 & \ddots \\ 0 & 0 & 1 & 0 &  \\  &  & \ddots &  & \end{pmatrix} , \label{eq:theta}
 \end{align}
where $\theta$ has mass dimension $-2$. Notice that $\mathcal S$ is stable by the product $\star$. Then, the Moyal product can be actually extended by duality to the following subalgebra of tempered distributions $\mathcal S'(\mathbb R^D)$ given by
\begin{align}
\mathcal M_\theta=\{T\in\mathcal S'(\mathbb R^D),\quad \forall f\in\mathcal S\quad T\star f\in\mathcal S\quad \text{and}\quad f\star T\in\mathcal S\}.
\end{align}

The Moyal space $(\mathcal M_\theta,\star,\dag)$ is a unital involutive associative algebra (where $\dag$ denotes the complex conjugation), and involves in particular the ``coordinate functions'' $x_\mu$, satisfying the following commutation relation: $[x_\mu,x_\nu]_\star =x_\mu\star x_\nu-x_\nu\star x_\mu=i\Theta_{\mu\nu}$. From this relation, and defining $\widetilde x_\mu=2\Theta^{-1}_{\mu\nu}x_\nu$, we deduce some useful properties of $\mathcal M_\theta$:
\begin{subequations}
\begin{align}
\forall f,h\in\mathcal M_\theta,\qquad &
\partial_\mu(f\star h)=\partial_\mu f\star h+f\star\partial_\mu h, \label{eq:relat1}\\
&\int d^4x\ f\star h=\int d^4x\ f.h,\label{eq:relat2}\\
&[\widetilde x_\mu,f]_\star=2i\partial_\mu f,\label{eq:relat3}\\
& \{\widetilde x_\mu,f\}_\star=\widetilde x_\mu\star f+f\star\widetilde x_\mu=2\widetilde x_\mu.f .\label{eq:relat4}
\end{align}
\end{subequations}

In the present noncommutative framework, the Yang-Mills theory can be built from real-valued gauge potentials $A_\mu$ defined on $\mathcal M_\theta$, stemming from the very definition of noncommutative connections. For more mathematical details, see \cite{CONNES,Dubois-Violette:1989vq,Masson:1999,Wallet:2007 em, de Goursac:2007gq}. Recall that the group of gauge transformations acts on the gauge potential as
\begin{align}
A_\mu^g &=g\star A_\mu\star g^\dag+ig\star\partial_\mu g^\dag\label{eq:gaugepot},
\end{align}
where $g\in\mathcal M_\theta$ is the gauge function and it satisfies $g^\dag\star g=g\star g^\dag=\mathbb I$. Recall also that there exists a special gauge potential in the Moyal space defined by $(-\frac{1}{2}\widetilde{x}_\mu)$. One can check that $(-\frac 12\widetilde x_\mu)^g=-\frac 12\widetilde x_\mu$ holds, stemming from the existence of a gauge invariant connection \cite{Dubois-Violette:1989vq,Masson:1999,Wallet:2007 em}, whose occurence is implied by the fact that all derivations on $\mathcal M_\theta$ are inner. From this gauge invariant potential, we construct the covariant coordinates \cite{Douglas:2001ba}:
\begin{align}
\mathcal A_\mu=A_\mu+\frac 12\widetilde x_\mu,\label{eq:covcoord}
\end{align}
which transform covariantly:
\begin{align}
\mathcal A^g_\mu=g\star\mathcal A_\mu\star g^\dag.\label{eq:gaugecov}
\end{align}
Then, the curvature $F_{\mu\nu}=\partial_\mu A_\nu-\partial_\nu A_\mu-i [A_\mu,A_\nu]_\star$, transforming as $F^g_{\mu\nu}=g\star F_{\mu\nu}\star g^\dag$, can be reexpressed in terms of $\mathcal A_\mu$ as
\begin{align}
F_{\mu\nu}=\Theta^{-1}_{\mu\nu}-i[\mathcal A_\mu,\mathcal A_\nu]_\star.\label{eq:Fcovcoord}
\end{align}
It is known that the naive noncommutative extension of the Yang-Mills action is plagued by UV/IR mixing \cite{Matusis:2000jf} which renders its renormalisability very unlikely, unless it is suitably modified. It turns out that the extension of the harmonic solution proposed in \cite{Grosse:2004yu} to the case of gauge theories has been achieved recently \cite{de Goursac:2007gq, Grosse:2007qx}. This singled out a class of potentially renormalisable theories on $\mathcal M_\theta$ whose action can be generically written as 
\begin{align}
S=\int d^Dx\ \Big(\frac{1}{4}F_{\mu\nu}\star F_{\mu\nu}+\frac{\Omega^2}{4}\{\mathcal{A}_\mu,\mathcal{A}_\nu\}_\star^2+\kappa \mathcal A_\mu\star\mathcal A_\mu\Big),\label{eq:actionYM}
\end{align}
in which $\Omega$ and $\kappa$ are real parameters, with mass dimensions respectively given by $[\Omega]=0$ and $[\kappa]=2$. This has been further shown to be related to a spectral triple \cite{Grosse:2007jy}.

\subsection{The matrix base.}

It will be convenient to represent elements on $\mathcal M_\theta$ with the help of the ``so-called'' matrix base. For more details, see \cite{Grosse:2003nw}. Recall that its elements $(b_{mn}^{(D)}(x))$ in $D$ dimensions are eigenfunctions of the harmonic oscillator hamiltonian $H=\frac{x^2}{2}$:
\begin{align}
H\star b_{mn}^{(D)}=\theta(|m|+\frac{1}{2})b_{mn}^{(D)} \qquad b_{mn}^{(D)}\star H=\theta(|n|+\frac{1}{2})b_{mn}^{(D)},
\end{align}
where $m,n\in\mathbb N^{\frac D2}$ and $|m|=\sum_{i=1}^{\frac D2}m_i$. In two dimensions, the expression of the elements $(b_{mn}^{(2)})=(f_{mn})$ of the matrix base in polar coordinates
\begin{align}
x_1=r\cos(\varphi),\qquad x_2=r\sin(\varphi),
\end{align}
is given by
\begin{align}
f_{mn}(x)=2(-1)^m\sqrt{\frac{m!}{n!}}e^{i(n-m)\varphi}\left(\frac{2r^2}{\theta}\right)^{\frac{n-m}{2}}L_m^{n-m}\left(\frac{2r^2}{\theta}\right) e^{-\frac{r^2}{\theta}},\label{eq:laguerre}
\end{align}
where the $L_n^k(x)$ are the associated Laguerre polynomials. The extension in four dimensions is straighforward. Namely, one has $m=(m_1,m_2)$, $n=(n_1,n_2)$ and
\begin{align}
b_{mn}^{(4)}(x)=f_{m_1,n_1}(x_1,x_2)\ f_{m_2,n_2}(x_3,x_4).\label{eq:basemat4}
\end{align}

The matrix base satisfies to the following properties:
\begin{align}
(b_{mn}^{(D)}\star b_{kl}^{(D)})(x)=\delta_{nk} b_{ml}^{(D)}(x),\\
\int d^Dx\ b_{mn}^{(D)}(x)=(2\pi\theta)^{\frac D2}\delta_{mn},\\
(b_{mn}^{(D)})^\dag(x)=b_{nm}^{(D)}(x).
\end{align}
We recall that this base defines therefore an isomorphism between the unital involutive Moyal algebra and a subalgebra of the unital involutive algebra of complex infinite-dimensional matrices. Indeed, for all $g\in\mathcal M_\theta$, there is a unique matrix $(g_{mn})$ satisfying
\begin{align}
\forall x\in\mathbb R^D\quad g(x)=\sum_{m,n\in\mathbb N^{\frac D2}}g_{mn}b_{mn}^{(D)}(x).
\end{align}
Then, this matrix is given by
\begin{align}
g_{mn}=\frac{1}{(2\pi\theta)^{\frac D2}}\int d^Dx\ g(x)b_{mn}^{(D)}(x).\label{eq:coeffmatrix}
\end{align}

\section{Symmetries of the vacua.}

In the following, we will have to solve the equation of motion. This is a difficult task. In this respect, it is convenient to exhibit some symmetries of the theory that will be used to constraint the expression for the solutions we look for. In fact, the group of symmetries of the euclidean Moyal algebra $\mathcal M_\theta$ in $D$ dimensions is
\begin{align}
G_D=SO(D)\cap Sp(D),
\end{align}
where $SO(D)$ is the group of rotations and $Sp(D)$ is the group of symplectic isomorphisms. $G_D$ acts on the field $A_\mu$ or $\mathcal A_\mu$ as
\begin{align}
\forall \Lambda\in G_D,\quad A^\Lambda_\mu(x)=\Lambda_{\mu\nu}A_\nu(\Lambda^{-1}x).
\end{align}
The action $S(\mathcal A)$ \eqref{eq:actionYM} is of course invariant under $G_D$. We further require that the new action $\widetilde S(\mathcal A_\mu^0,\delta \mathcal A_\mu)=S(\mathcal A_\mu^0+\delta\mathcal A_\mu)$, obtained from the expansion of $S$ around a non-trivial vacuum $\mathcal A_\mu^0$, is also invariant under $G_D$. This means that $\widetilde S(\mathcal A_\mu^0,\delta\mathcal A^\Lambda_\mu)=\widetilde S(\mathcal A_\mu^0,\delta\mathcal A_\mu)$, where $\Lambda\in G_D$ do not affect the vacuum $\mathcal A^0_\mu$. Since $S(\mathcal A_\mu)$ is invariant under $G_D$, this is equivalent to
\begin{align}
\forall \Lambda\in G_D,\quad \widetilde S((\mathcal A_\mu^0)^\Lambda,\delta\mathcal A_\mu)=\widetilde S(\mathcal A_\mu^0,\delta\mathcal A_\mu).\label{eq:condvac}
\end{align}
This last relation implies that the vacuum is invariant under $G_D$. Indeed, upon using the expression \eqref{eq:actionYM2} of the action given in the next section, the part of \eqref{eq:condvac} quadratic in $\delta\mathcal A_\mu$ can be written as
\begin{align}
&\int d^Dx\ \Big(-\frac{(1-\Omega^2)}{2}(2\mathcal{A}^0_\mu\star\mathcal{A}^0_\nu\star\delta\mathcal{A}_\mu\star \delta\mathcal A_\nu+2\mathcal{A}^0_\mu\star\mathcal{A}^0_\nu\star\delta\mathcal{A}_\nu\star \delta\mathcal A_\mu\nonumber\\
&+2\mathcal{A}^0_\mu\star\delta\mathcal{A}_\nu\star\mathcal{A}^0_\mu\star \delta\mathcal A_\nu)+ \frac{(1+\Omega^2)}{2}(2\mathcal{A}^0_\mu\star\mathcal{A}^0_\mu\star\delta\mathcal{A}_\nu\star\delta \mathcal A_\nu+2\mathcal{A}^0_\mu\star\mathcal{A}^0_\nu\star\delta\mathcal{A}_\nu\star \delta\mathcal A_\mu\nonumber\\
&+\mathcal{A}^0_\mu\star\delta\mathcal{A}_\mu\star\mathcal{A}^0_\nu\star \delta\mathcal A_\nu+\delta\mathcal{A}_\mu\star\mathcal{A}^0_\mu\star\delta\mathcal{A}_\nu\star \mathcal A^0_\nu) \Big) =\nonumber\\
&\int d^Dx\ \Big(-\frac{(1-\Omega^2)}{2}(2(\mathcal{A}^0_\mu)^\Lambda\star(\mathcal{A}^0_\nu)^\Lambda\star\delta\mathcal{A}_\mu\star \delta\mathcal A_\nu+2(\mathcal{A}^0)^\Lambda_\mu\star(\mathcal{A}^0_\nu)^\Lambda\star\delta\mathcal{A}_\nu\star \delta\mathcal A_\mu\nonumber\\
&+2(\mathcal{A}^0_\mu)^\Lambda\star\delta\mathcal{A}_\nu\star(\mathcal{A}^0_\mu)^\Lambda\star \delta\mathcal A_\nu)+ \frac{(1+\Omega^2)}{2}(2(\mathcal{A}^0_\mu)^\Lambda\star(\mathcal{A}^0_\mu)^\Lambda\star\delta\mathcal{A}_\nu\star\delta \mathcal A_\nu\nonumber\\
&+2(\mathcal{A}^0_\mu)^\Lambda\star(\mathcal{A}^0_\nu)^\Lambda\star\delta\mathcal{A}_\nu\star \delta\mathcal A_\mu+(\mathcal{A}^0_\mu)^\Lambda\star\delta\mathcal{A}_\mu\star(\mathcal{A}^0_\nu)^\Lambda\star \delta\mathcal A_\nu+\delta\mathcal{A}_\mu\star(\mathcal{A}^0_\mu)^\Lambda\star\delta\mathcal{A}_\nu\star (\mathcal A^0_\nu)^\Lambda) \Big).\label{eq:bigexp}
\end{align}
This relation is true for all the fluctuations $\delta\mathcal A_\mu$, so it holds at the level of the lagrangians involved in the integrals. Assuming now $\delta\mathcal A_\mu(x)=\delta_{\mu\rho}$, for some fixed $\rho$, we obtain from \eqref{eq:bigexp}
\begin{align}
2\Omega^2(\mathcal{A}^0_\mu\mathcal{A}^0_\mu+2(\mathcal{A}^0_\rho)^2-(\mathcal{A}^0_\mu)^\Lambda(\mathcal{A}^0_\mu)^\Lambda-2((\mathcal{A}^0_\rho)^\Lambda)^2)=0,
\end{align}
where the index $\rho$ is not summed over. It is now easy to get $((\mathcal{A}^0_\rho)^\Lambda)^2=(\mathcal{A}^0_\rho)^2$, and since $G_D$ is a connected Lie group,
\begin{align}
(\mathcal{A}^0_\rho)^\Lambda(x)=\mathcal{A}^0_\rho(x).
\end{align}

The vacuum is invariant under $G_D$, i.e., dropping from now on in $\mathcal A^0_\mu$ the superscript 0, $\mathcal A_\mu(x)=\Lambda_{\mu\nu}\mathcal A_\nu(\Lambda^{-1}x)$ for all $\Lambda\in G_D$.  Since the identity matrix and $2\Theta^{-1}$ are the only matrices up to a scalar multiplication that commute with $G_D$, we can write that
\begin{align}
\mathcal A_\mu(x)=\phi_1(x)x_\mu+\phi_2(x)2\Theta^{-1}_{\mu\nu}x_\nu= \phi_1(x)x_\mu+\phi_2(x)\widetilde x_\mu,
\end{align}
where $\phi_1$ and $\phi_2$ are two scalar fields invariant under $G_D$. Then, $G_D$ is isomorphic to $U(\frac D2)$ and the isomorphism is described by associating to each coefficient $u_{ij}$ of a matrix of $U(\frac D2)$, the submatrix
\begin{align}
\begin{pmatrix} \text{Re}(u_{ij}) & -\text{Im}(u_{ij}) \\ \text{Im}(u_{ij}) & \text{Re}(u_{ij})  \end{pmatrix},
\end{align}
in the place $(i,j)$ of the matrix of $G_D$. From the theory of the invariants of $U(\frac D2)$ \cite{Weyl}, one infers that $\phi_1$ and $\phi_2$ are therefore functions only on $x^2$. Then, the general expression for $\mathcal A_\mu$ can be written as
\begin{align}
\mathcal A_\mu(x)=\Phi_1(x^2)x_\mu+\Phi_2(x^2)\widetilde x_\mu.\label{eq:solform}
\end{align}
This form will be extensively used to solve the equation of motion in the following section.

\section{Solving the equation of motion.}

By using the definition of the covariant coordinate \eqref{eq:covcoord}, namely $\mathcal A_\mu=A_\mu+\frac 12\widetilde x_\mu$, the action \eqref{eq:actionYM} can be rewritten as
\begin{align}
S=\int d^Dx\ \Big(-\frac{(1-\Omega^2)}{2}\mathcal{A}_\mu\star\mathcal{A}_\nu\star\mathcal{A}_\mu\star \mathcal A_\nu + \frac{(1+\Omega^2)}{2}\mathcal{A}_\mu\star\mathcal{A}_\mu\star\mathcal{A}_\nu\star \mathcal A_\nu +\kappa\mathcal{A}_\mu\star \mathcal A_\mu\Big),\label{eq:actionYM2}
\end{align}
Then, the corresponding equation of motion $\frac{\delta S}{\delta \mathcal A_\mu(x)}=0$ is given by
\begin{align}
-2(1-\Omega^2)\mathcal{A}_\nu\star\mathcal{A}_\mu\star\mathcal{A}_\nu + (1+\Omega^2)\mathcal{A}_\mu\star\mathcal{A}_\nu\star\mathcal{A}_\nu + (1+\Omega^2)\mathcal{A}_\nu\star\mathcal{A}_\nu\star\mathcal{A}_\mu + 2\kappa\mathcal{A}_\mu=0.\label{eq:motion}
\end{align}
Due to the very structure of the Moyal product, this is a complicated integro-differential equation for which no known algorithm to solve it does exist so far. Notice that it supports the trivial solution $\mathcal A_\mu(x)=0$, which, however, is not so interesting since expanding the action around it give rise to a non-dynamical matrix model, as already noted in \cite{deGoursac:2007qi}. It turns out that \eqref{eq:motion} supports other nontrivial solutions. These can be conveniently determined for $D=2$ and $D=4$ using the matrix base \eqref{eq:laguerre} and \eqref{eq:basemat4} as we now show in the rest of this section.

\subsection{The case D=2.}

When $D=2$, it is convenient to define
\begin{align}
Z(x)=\frac{\mathcal A_1(x)+i\mathcal A_2(x)}{\sqrt{2}}\qquad Z^\dag(x)=\frac{\mathcal A_1(x)-i\mathcal A_2(x)}{\sqrt{2}}.\label{eq:coordcomp}
\end{align}
Then, the action can be expressed as
\begin{align}
S=\int d^2x\Big((-1+3\Omega^2)Z\star Z\star Z^\dag\star Z^\dag+(1+\Omega^2)Z\star Z^\dag\star Z\star Z^\dag+2\kappa Z\star Z^\dag\Big),\label{eq:actmatrix}
\end{align}
so that the equation of motion takes the form
\begin{align}
(3\Omega^2-1)(Z^\dag\star Z\star Z+Z\star Z\star Z^\dag)+2(1+\Omega^2)Z\star Z^\dag\star Z+2\kappa Z=0.\label{eq:fondam}
\end{align}
Expressing now $Z(x)$ in the matrix base, namely
\begin{align}
Z(x)=\sum_{m,n=0}^\infty Z_{mn} f_{mn}(x),
\end{align}
\eqref{eq:fondam} becomes a cubic infinite-dimensional matrix equation. In view of the discussion carried out in section 3, we now look for the symmetric solutions of the form given by \eqref{eq:solform}, namely
\begin{align}
Z(x)=\Phi_1(x^2)\left(\frac{x_1+ix_2}{\sqrt 2}\right)+\Phi_2(x^2)\left(\frac{\widetilde x_1+i\widetilde x_2}{\sqrt 2}\right).\label{eq:ansatz2}
\end{align}
To translate \eqref{eq:ansatz2} into the matrix base, we first note that the expression of the matrix coefficients of $Z(x)$ is given by \eqref{eq:coeffmatrix}
\begin{align}
Z_{mn}=\frac{1}{2\pi\theta}\int d^2x Z(x)f_{nm}(x).
\end{align}
In polar coordinates $(r,\varphi)$, we have
\begin{align}
x_1+ix_2=re^{i\varphi},\qquad \widetilde x_1+i\widetilde x_2=-\frac{2i}{\theta}re^{i\varphi},
\end{align}
and
\begin{align}
Z_{mn}=\frac{(-1)^n}{2\pi\theta\sqrt 2}\sqrt{\frac{n!}{m!}}\int rdrd\varphi e^{i(m-n)\varphi}\left(\frac{2r^2}{\theta}\right)^{\frac{m-n}{2}}L_n^{m-n}\left(\frac{2r^2}{\theta}\right) e^{-\frac{r^2}{\theta}}(\Phi_1(r^2)re^{i\varphi}-\Phi_2(r^2)\frac{2i}{\theta}re^{i\varphi}).
\end{align}
By performing the integration over $\varphi$, we easily find that
\begin{align}
Z_{mn}=\frac{(-1)^n}{\theta\sqrt 2}\sqrt{\frac{n!}{m!}}\int r^2dr \left(\frac{2r^2}{\theta}\right)^{\frac{m-n}{2}}L_n^{m-n}\left(\frac{2r^2}{\theta}\right) e^{-\frac{r^2}{\theta}}(\Phi_1(r^2)-\Phi_2(r^2)\frac{2i}{\theta})\delta_{m+1,n}.
\end{align}
Then, defining $z=\frac{2r^2}{\theta}$ and
\begin{align}
a_m=\frac{(-1)^{m+1}}{4}\sqrt{\frac{m+1}{\theta}}\int dz L_{m+1}^{-1}(z)e^{-\frac z2}(\Phi_2(\frac{\theta z}{2})+\frac{i\theta}{2}\Phi_1(\frac{\theta z}{2})),
\end{align}
we obtain
\begin{align}
Z_{mn}=-ia_m\delta_{m+1,n}.\label{eq:ansatz}
\end{align}
This, inserted into \eqref{eq:fondam} yields
\begin{align}
\forall m\in\mathbb N,\quad a_m\left((3\Omega^2-1)(|a_{m-1}|^2+|a_{m+1}|^2)+2(1+\Omega^2)|a_m|^2+2\kappa\right)=0,\label{eq:mvtmatr}
\end{align}
where it is understood that $a_{-1}=0$. Then, \eqref{eq:mvtmatr} implies
\begin{align}
\forall m\in\mathbb N,\quad (i)\ a_m=0\quad\text{or}\quad (ii)\ (3\Omega^2-1)(|a_{m-1}|^2+|a_{m+1}|^2) +2(1+\Omega^2)|a_m|^2+2\kappa=0.
\end{align}
From now on, we will focus only on the second condition $(ii)$ (this will be more closely discussed in section 6, see hypothesis $(\mathfrak H)$):
\begin{align}
\forall m\in\mathbb N,\quad (3\Omega^2-1)(|a_{m-1}|^2+|a_{m+1}|^2)+2(1+\Omega^2)|a_m|^2+2\kappa=0.\label{eq:princ1}
\end{align}
Upon setting $u_{m+1}=|a_m|^2$, \eqref{eq:princ1} becomes
\begin{align}
\forall m\in\mathbb N,\quad (3\Omega^2-1)(u_m+u_{m+2})+2(1+\Omega^2)u_{m+1}+2\kappa=0,\label{eq:princ2}
\end{align}
which is a non homogenous linear iterative equation of second order with boundary condition $u_0=0$. This will be solved in subsection 5.1.

\subsection{The case D=4.}

The case $D=4$ can be straighforwardly adapted from the two-dimensional case. Owing to the fact that two symplectic pairs are now involved in the four-dimensional Moyal space, we define two complex quantities, namely
\begin{align}
Z_1(x)=\frac{\mathcal A_1(x)+i\mathcal A_2(x)}{\sqrt{2}}\qquad Z_2(x)=\frac{\mathcal A_3(x)+i\mathcal A_4(x)}{\sqrt{2}}.
\end{align}
Then, using these new variables, equation \eqref{eq:actionYM2} can be conveniently reexpressed as
\begin{align}
S=&\int d^4x \Big( (-1+3\Omega^2)Z_1\star Z_1\star Z_1^\dag\star Z_1^\dag+(1+\Omega^2)Z_1\star Z_1^\dag\star Z_1\star Z_1^\dag+2\kappa Z_1\star Z_1^\dag\nonumber\\
&+(-1+3\Omega^2)Z_2\star Z_2\star Z_2^\dag\star Z_2^\dag+(1+\Omega^2)Z_2\star Z_2^\dag\star Z_2\star Z_2^\dag+2\kappa Z_2\star Z_2^\dag\nonumber\\
&-2(1-\Omega^2)Z_1\star Z_2\star Z_1^\dag\star Z_2^\dag -2(1-\Omega^2)Z_2\star Z_1\star Z_2^\dag\star Z_1^\dag +(1+\Omega^2)Z_1\star Z_1^\dag\star Z_2\star Z_2^\dag\nonumber\\
&+(1+\Omega^2)Z_2\star Z_1\star Z_1^\dag\star Z_2^\dag +(1+\Omega^2)Z_1\star Z_2\star Z_2^\dag\star Z_1^\dag +(1+\Omega^2)Z_1^\dag\star Z_1\star Z_2^\dag\star Z_2\Big).\label{eq:actmatrix4d}
\end{align}
From \eqref{eq:actmatrix4d}, we derive the equations of motion
\begin{subequations}
\begin{align}
&(3\Omega^2-1)(Z_1^\dag\star Z_1\star Z_1+Z_1\star Z_1\star Z_1^\dag)+(1+\Omega^2)(2Z_1\star Z_1^\dag\star Z_1+Z_2\star Z_2^\dag\star Z_1\nonumber\\
&+Z_2^\dag \star Z_2\star Z_1+Z_1\star Z_2\star Z_2^\dag+Z_1\star Z_2^\dag\star Z_2) -2(1-\Omega^2)(Z_2^\dag\star Z_1\star Z_2+Z_2\star Z_1\star Z_2^\dag)\nonumber\\
&+2\kappa Z_1=0,\label{eq:fondam4a}\\
&(3\Omega^2-1)(Z_2^\dag\star Z_2\star Z_2+Z_2\star Z_2\star Z_2^\dag)+(1+\Omega^2)(2Z_2\star Z_2^\dag\star Z_2+Z_1\star Z_1^\dag\star Z_2\nonumber\\
&+Z_1^\dag \star Z_1\star Z_2+Z_2\star Z_1\star Z_1^\dag+Z_2\star Z_1^\dag\star Z_1) -2(1-\Omega^2)(Z_1^\dag\star Z_2\star Z_1+Z_1\star Z_2\star Z_1^\dag)\nonumber\\
&+2\kappa Z_2=0.\label{eq:fondam4b}
\end{align}
\end{subequations}
Notice that \eqref{eq:fondam4a} and \eqref{eq:fondam4b} are exchanged upon performing the exchange $Z_1\rightleftarrows Z_2$. Now we again specialize to the symmetric solutions of the form \eqref{eq:solform}, namely
\begin{align}
Z_1(x)=\frac{1}{\sqrt 2}(\Phi_1(x^2)(x_1+ix_2)+\Phi_2(x^2)(\widetilde x_1+i\widetilde x_2)),\nonumber\\
Z_2(x)=\frac{1}{\sqrt 2}(\Phi_1(x^2)(x_3+ix_4)+\Phi_2(x^2)(\widetilde x_3+i\widetilde x_4)).
\end{align}
Then, in view of \eqref{eq:basemat4} and \eqref{eq:coeffmatrix}, one has
\begin{align}
(Z_1)_{m,n}=\frac{1}{(2\pi\theta)^2}\int d^4x\ Z_1(x)f_{n_1,m_1}(x_1,x_2)f_{n_2,m_2}(x_3,x_4).
\end{align}
Let us introduce the polar coordinates associated to each symplectic pair:
\begin{align}
x_1=r_1\cos(\varphi_1)&\qquad x_2=r_1\sin(\varphi_1)\nonumber\\
x_3=r_2\cos(\varphi_2)&\qquad x_4=r_2\sin(\varphi_2)\nonumber\\
r^2=&r_1^2+r_2^2.\label{eq:polcoord4d}
\end{align}
Then, by using \eqref{eq:laguerre} and integrating over the two angular variables $\varphi_1$ and $\varphi_2$, one obtains
\begin{align}
(Z_1)_{m,n}=&\frac{2(-1)^{m_1+m_2+1}}{\theta}\sqrt{\frac{m_1+1}{\theta}}\int r_1dr_1r_2dr_2(\Phi_1(r^2)-\frac{2i}{\theta}\Phi_2(r^2))\nonumber\\
&e^{-\frac{r^2}{\theta}}L_{m_1+1}^{-1}(\frac{2r^2}{\theta})L_{m_2}^0(\frac{2r^2}{\theta})\delta_{m_1+1,n_1}\delta_{m_2,n_2}.\label{eq:calcinter}
\end{align}
Let us now integrate by part this expression. Defining $z_1=\frac{2r_1^2}{\theta}$, $z_2=\frac{2r_2^2}{\theta}$, $z=z_1+z_2$, and denoting by $F(z)$ one primitive function of $(\Phi_2(\frac{\theta z}{2})+\frac{i\theta}{2}\Phi_1(\frac{\theta z}{2}))e^{-\frac z2}$, \eqref{eq:calcinter} leads to
\begin{align}
(Z_1)_{m,n}=-i\frac{(-1)^{m_1+m_2+1}}{4}\sqrt{\frac{m_1+1}{\theta}}\int dz_1dz_2\ F(z)L_{m_1}^0(z_1)L_{m_2}^0(z_2)\delta_{m_1+1,n_1}\delta_{m_2,n_2},
\end{align}
where we have used
\begin{align}
\frac{d}{dx}L_{m_1+1}^{-1}(x)=-L_{m_1}^0(x).
\end{align}
A similar derivation holds for $(Z_2)_{m,n}$. Finally, using the symmetry argument developped in section 3, we find
\begin{align}
(Z_1)_{m,n}&=-ia_{m_1m_2}\sqrt{m_1+1}\delta_{m_1+1,n_1}\delta_{m_2,n_2},\\
(Z_2)_{m,n}&=-ia_{m_1m_2}\sqrt{m_2+1}\delta_{m_1,n_1}\delta_{m_2+1,n_2},
\end{align}
where $a_{m_1,m_2}\in\mathbb C$ is symmetric upon the exchange of $m_1$ and $m_2$. Then, \eqref{eq:fondam4a} and \eqref{eq:fondam4b} become
\begin{subequations}
\label{eq:fondam5}
\begin{align}
&(3\Omega^2-1)(m_1|a_{m_1-1,m_2}|^2+(m_1+2)|a_{m_1+1,m_2}|^2)a_{m_1,m_2} +(1+\Omega^2)(2(m_1+1)|a_{m_1,m_2}|^2\nonumber\\
&+(m_2+1)|a_{m_1,m_2}|^2+m_2|a_{m_1,m_2-1}|^2+(m_2+1)|a_{m_1+1,m_2}|^2+m_2|a_{m_1+1,m_2-1}|^2)a_{m_1,m_2}\nonumber\\
&-2(1-\Omega^2)(m_2|a_{m_1,m_2-1}|^2a_{m_1+1,m_2-1}+(m_2+1)\overline a_{m_1+1,m_2}a_{m_1,m_2+1}a_{m_1,m_2}) +2\kappa a_{m_1,m_2}=0,\label{eq:fondam5a}\\
&(3\Omega^2-1)(m_2|a_{m_1,m_2-1}|^2+(m_2+2)|a_{m_1,m_2+1}|^2)a_{m_1,m_2} +(1+\Omega^2)(2(m_2+1)|a_{m_1,m_2}|^2\nonumber\\
&+(m_1+1)|a_{m_1,m_2}|^2+m_1|a_{m_1-1,m_2}|^2+(m_1+1)|a_{m_1,m_2+1}|^2+m_1|a_{m_1-1,m_2+1}|^2)a_{m_1,m_2}\nonumber\\
&-2(1-\Omega^2)(m_1|a_{m_1-1,m_2}|^2a_{m_1-1,m_2+1}+(m_1+1)\overline a_{m_1,m_2+1}a_{m_1+1,m_2}a_{m_1,m_2}) +2\kappa a_{m_1,m_2}=0.\label{eq:fondam5b}
\end{align}
\end{subequations}
As we did for the case $D=2$, we assume now that $a_{m_1,m_2}\neq0$ (for a more detailled discussion of this point, see section 6. This will be called hypothesis $(\mathfrak H)$). Combining this latter assumption with \eqref{eq:fondam5}, it can be shown that $a_{m_1,m_2}$ depends only on $m=m_1+m_2$. The corresponding proof is presented in the appendix A. Now, if we define $v_{m_1+m_2+1}=|a_{m_1,m_2}|^2$, \eqref{eq:fondam5} is equivalent to
\begin{align}
\forall m\in\mathbb N,\quad (3\Omega^2-1)(mv_m+(m+3)v_{m+2})+(1+\Omega^2)(2m+3)v_{m+1}+2\kappa=0,\label{eq:princ4}
\end{align}
which is a non homogenous linear iterative equation of second order with non constant coefficients, with boundary condition $v_0=0$. Notice also that this equation is very close to this of the case $D=2$, defining $v_m=\frac{u_m}{m}$ in \eqref{eq:princ2}.

\section{Solutions.}

In this section, we solve the equations \eqref{eq:princ2} and \eqref{eq:princ4} to obtain the vacuum configurations for $D=2$ and $D=4$.

\subsection{The case D=2.}

Let us consider the equation \eqref{eq:princ2}
\begin{align}
\forall m\in\mathbb N,\quad (3\Omega^2-1) &(u_m+u_{m+2})+2(1+\Omega^2)u_{m+1}+2\kappa=0,\nonumber\\
&u_0=0\quad\text{and}\quad u_m\geq0\label{eq:princ25}.
\end{align}
For $\Omega^2\neq \frac 13$, we define
\begin{align}
r=\frac{1+\Omega^2+\sqrt{8\Omega^2(1-\Omega^2)}}{1-3\Omega^2},
\end{align}
and one has $\frac 1r=\frac{1+\Omega^2-\sqrt{8\Omega^2(1-\Omega^2)}}{1-3\Omega^2}$. Then, it is easy to realise that \eqref{eq:princ25} supports different types of solutions according to the range for the values taken by $\Omega$. Namely, one has
\begin{itemize}
\item for $\Omega^2=0$, $\kappa=0$, $u_m=\alpha m$ and $\alpha\geq0$;
\item for $0<\Omega^2<\frac 13$, $u_m=\alpha(r^{m}-r^{-m})-\frac{\kappa}{4\Omega^2}(1-r^{-m})$, $\alpha\geq0$ and $r>1$;
\item for $\Omega^2=\frac 13$, $\kappa\leq0$ and $u_m=-\frac{3\kappa}{4}$;
\item for $\frac 13<\Omega^2<1$, $\kappa\leq0$, $u_m=-\frac{\kappa}{4\Omega^2}(1-r^{-m})$ and $r<-1$;
\item for $\Omega^2=1$, $\kappa\leq0$ and $u_m=-\frac{\kappa}{4}(1-(-1)^{-m})$.
\end{itemize}
Notice that the solution for $\Omega=0$ corresponds to the commutative case: $u_m=\frac{m}{\theta}$ is equivalent to a vacuum $\mathcal A_\mu(x)=\frac 12\widetilde x_\mu$ or $A_\mu=0$. Then, it is possible to choose $\alpha$ depending on $\Omega$ so that the solution is continous in $\Omega$ near 0. With the following Taylor expansions
\begin{align}
&r^{m}-r^{-m}=4\sqrt 2m\Omega+O(\Omega^3),\nonumber\\
&1-r^{-m}=2\sqrt 2m\Omega-4m^2\Omega^2+O(\Omega^3),\nonumber\\
&\alpha(\Omega)=\frac{1}{\Omega}(\alpha_0+O(\Omega)),\nonumber\\
\end{align}
one deduces that $\kappa$ must have the same asymptotic behaviour as $\Omega$ near 0. If $\kappa=\kappa_0\Omega+O(\Omega)$,
\begin{align}
u_m=4\sqrt 2m\alpha_0-\frac{\sqrt 2m}{2}\kappa_0+O(\Omega).
\end{align}
For $\alpha_0=\frac{\kappa_0}{8}+\frac{4}{4\theta\sqrt 2}$, we find the commutative limit for the vacuum: $A_\mu=0$, and the gauge potential is massless ($\lim_{\Omega\to0}\kappa=0$).

Consider now the asymptotic behaviour of the vacuum in the configuration space for $x^2\to\infty$. If $0<\Omega^2<\frac 13$ and $\alpha\neq0$, then $\sqrt{u_m}\thicksim_{m\to\infty}r^{\frac m2}$ and $r>1$. As a consequence \cite{Gracia-Bondia:1987kw}, the solution $\mathcal A_\mu(x)$ of the equation of motion does not belong to the Moyal algebra. So we require that $\alpha=0$. Then, for $\Omega\neq0$, $\kappa$ has to be negative and $u_m$ has a finite limit. This indicates that $\mathcal A_\mu(x)$ has a constant limit in $x^2\to\infty$.

Let us try to obtain an expression for the vacuum in the configuration space. Using the variable \eqref{eq:coordcomp}, the solution is
\begin{align}
Z(x)&=\sum_{m,n=0}^\infty -ia_m\delta_{m+1,n}f_{m,n}(x)\nonumber\\
&=-i\sum_{m=0}^\infty a_mf_{m,m+1}(x),
\end{align}
with $a_m=e^{i\xi_m}\sqrt{u_m}$ and $\xi_m\in\mathbb R$ an arbitrary phase. Using the equation \eqref{eq:laguerre}, we obtain
\begin{align}
Z(x)=-2i\sqrt z e^{\frac z2}e^{i\varphi}\sum_{m=0}^\infty \frac{(-1)^m}{\sqrt{m+1}}a_mL^1_m(z),
\end{align}
where $z=\frac{2r^2}{\theta}$. Then, the use of the following property
\begin{align}
L_m^k(z)=\frac{e^zz^{-\frac k2}}{m!}\int_0^\infty dt\ e^{-t}t^{m+\frac k2}J_k(2\sqrt{tz})\label{eq:laguerreint}
\end{align}
permits one to express $Z(x)$ as
\begin{align}
Z(x)=-2ie^{\frac z2}e^{i\varphi}\int_0^\infty dt\ e^{-t}\sqrt tJ_1(2\sqrt{tz})\sum_{m=0}^\infty \frac{(-1)^ma_m}{m!\sqrt{m+1}}t^m,
\end{align}
with $J_k(x)$ is the $k$-th function of Bessel of the first kind. Since $-i\sqrt{\frac{2z}{\theta}}e^{i(\varphi+\xi_m)}=(\widetilde x_1+i\widetilde x_2)\cos(\xi_m)+\frac{2}{\theta}(x_1+ix_2)\sin(\xi_m)$, one can deduce the expression of the vacuum in two dimensions
\begin{align}
\mathcal A_\mu(x)=2\sqrt{\theta}\frac{e^{\frac z2}}{\sqrt z}\int_0^\infty dt\ e^{-t}\sqrt tJ_1(2\sqrt{tz}) \sum_{m=0}^\infty \frac{(-1)^m\sqrt{u_{m+1}}}{m!\sqrt{m+1}}t^m\left(\widetilde x_\mu\cos(\xi_m)+\frac{2}{\theta}x_\mu\sin(\xi_m)\right),\label{eq:vacx2}
\end{align}
with $u_m$ given above.

\subsection{A special case.}

In this subsection, we consider the case $\Omega^2=\frac 13$ and $\kappa<0$ in two dimensions. We have found in the previous subsection that the solution is given by $u_m=-\frac{3\kappa}{4}$ and $a_m=e^{i\xi_m}\sqrt{u_m}$. Set now $\xi_m=0$ and let us check that this solution of the equation of motion is a minimum of the action. The action can be expanded around the vacuum $Z(x)$ for these values of parameters and its quadratic part is given by
\begin{align}
\widetilde S_{quadr}=\int d^2x\Big(& 2\kappa \delta Z\star\delta Z^\dag+4Z\star Z^\dag\star\delta Z\star\delta Z^\dag+4Z^\dag\star Z\star \delta Z^\dag\star\delta Z \nonumber\\
&+2Z\star\delta Z^\dag\star Z\star\delta Z^\dag+2Z^\dag\star\delta Z\star Z^\dag\star\delta Z\Big),\label{eq:actquad}
\end{align}
where $\delta Z(x)$ is the fluctuation. Then, denoting $\alpha=-\pi\theta\kappa>0$, \eqref{eq:actquad} can be reexpressed in the matrix base as
\begin{align}
\widetilde S_{quadr}=4\alpha \delta Z_{m,n}\delta Z^\dag_{n,m}-2\alpha\delta Z^\dag_{m+1,n}\delta Z^\dag_{n+1,m} -2\alpha \delta Z_{m,n+1}\delta Z_{n,m+1}.
\end{align}
Using \eqref{eq:coordcomp}, one can find that
\begin{align}
\widetilde S_{quadr}=&2\alpha (\delta\mathcal A_1)_{m,n}(\delta\mathcal A_1)_{n,m}-\alpha (\delta\mathcal A_1)_{m,n}(\delta\mathcal A_1)_{n+1,m-1} -\alpha(\delta\mathcal A_1)_{m,n}(\delta\mathcal A_1)_{n-1,m+1}\nonumber\\
&+2\alpha (\delta\mathcal A_2)_{m,n}(\delta\mathcal A_2)_{n,m}+\alpha (\delta\mathcal A_2)_{m,n}(\delta\mathcal A_2)_{n+1,m-1} +\alpha(\delta\mathcal A_2)_{m,n}(\delta\mathcal A_2)_{n-1,m+1}\nonumber\\
&+2i\alpha(\delta\mathcal A_1)_{m,n} (\delta\mathcal A_2)_{n+1,m-1}-2i\alpha(\delta\mathcal A_1)_{m,n}(\delta\mathcal A_2)_{n-1,m+1}.
\end{align}
By defining the variable
\begin{align}
X_{m,n}={\begin{pmatrix} (\delta\mathcal A_1)_{m,n} \\ (\delta\mathcal A_2)_{m,n} \end{pmatrix}}\label{eq:defx},
\end{align}
we find the following expression for \eqref{eq:actquad}
\begin{align}
\widetilde S_{quadr}=2\alpha X_{m,n}^T {\begin{pmatrix} 1&0 \\ 0&1 \end{pmatrix}}X_{n,m}+\alpha X_{m,n}^T {\begin{pmatrix} -1&i \\ i&1 \end{pmatrix}} X_{n+1,m-1}+\alpha X_{m,n}^T {\begin{pmatrix} -1&-i \\ -i&1 \end{pmatrix}}X_{n-1,m+1}.\label{eq:actquad2}
\end{align}
The operator involved in \eqref{eq:actquad2} is
\begin{align}
G_{m,n;k,l}=2\alpha {\begin{pmatrix} 1&0 \\ 0&1 \end{pmatrix}}\delta_{n,k}\delta_{m,l}+\alpha {\begin{pmatrix} -1&i \\ i&1 \end{pmatrix}}\delta_{n+1,k}\delta_{m,l+1}+\alpha {\begin{pmatrix} -1&-i \\ -i&1 \end{pmatrix}}\delta_{n,k+1}\delta_{m+1,l}.\label{eq:propagexp}
\end{align}
The above solution is a minimum for the action provided \eqref{eq:propagexp} is a positive operator. This can be shown indeed once it is realised that $G_{m,n;k,l}$ depends actually only on two indices, since the following identity among indices $m+n=k+l$ holds here. It follows that $G_{m,\gamma-m;\gamma-l,l}$ with $\gamma=m+n=k+l$ does not depend on $\gamma$ and therefore $G_{m,\gamma-m;\gamma-l,l}=G_{m,l}$. Then, one has
\begin{align}
G_{ml}=2\alpha {\begin{pmatrix} 1&0 \\ 0&1 \end{pmatrix}}\delta_{m,l}+\alpha {\begin{pmatrix} -1&i \\ i&1 \end{pmatrix}}\delta_{m,l+1}+\alpha {\begin{pmatrix} -1&-i \\ -i&1 \end{pmatrix}}\delta_{m+1,l}.
\end{align}
This operator can be represented by an infinite-dimensional matrix. Let us set a cut-off $N$ on the dimension of this matrix: $m,l\leq N$.
\begin{align}
G^{(N)}=\alpha {\begin{pmatrix} 2&0&-1&-i&0&0& \\ 0&2&-i&1&0&0& \\ -1&i&2&0&-1&-i&\ldots \\ i&1&0&2&-i&1& \\ 0&0&-1&i&2&0& \\ 0&0&i&1&0&2& \\ & & &\vdots& & &\ddots \end{pmatrix}}
\end{align}
is now a $2N\times 2N$ matrix, which is diagonalisable. Indeed:
\begin{itemize}
\item $2\alpha$ is a two-fold degenerate eigenvalue, with $(i,1,0,\dots,0)$ and $(0,\dots,0,-i,1)$ as associated eigenvectors.
\item 0 is a $(N-1)$-fold degenerate eigenvalue, with $(i,-1,i,1,0,\dots,0)$, $(0,0,i,-1,i,1,0,\dots,0)$, ..., $(0,\dots,0,i,-1,i,1,0,\dots,0)$, ..., $(0,\dots,0,i,-1,i,1)$ as associated eigenvectors.
\item $4\alpha$ is a $(N-1)$-fold degenerate eigenvalue, with $(-i,1,i,1,0,\dots,0)$, $(0,0,-i,1,i,1,0,\dots,0)$, ..., $(0,\dots,0,-i,1,i,1,0,\dots,0)$, ..., $(0,\dots,0,-i,1,i,1)$ as associated eigenvectors.
\end{itemize}
Since $\alpha=-\pi\theta\kappa>0$, $Z_{mn}=-i\sqrt{-\frac{3\kappa}{4}}\delta_{m+1,n}$, or equivalently
\begin{align}
\mathcal A_\mu(x)=\sqrt{-3\kappa\theta}\left(\frac{e^{\frac z2}}{\sqrt z}\int_0^\infty dt\ e^{-t}\sqrt tJ_1(2\sqrt{tz}) \sum_{m=0}^\infty \frac{(-1)^mt^m}{m!\sqrt{m+1}}\right)\widetilde x_\mu,
\end{align}
is a degenerated minimum of the action \eqref{eq:actionYM} for $\Omega^2=\frac 13$, $\kappa<0$, and where $z=\frac{2x^2}{\theta}$.

\subsection{The case D=4.}

The equation \eqref{eq:princ4} looks like \eqref{eq:princ2}, but the non triviality of the coefficients of this linear iterative equation makes it much more difficult to solve. Let us introduce the following auxiliary function
\begin{align}
y(x)=\sum_{m=1}^\infty v_mx^m,
\end{align}
since $v_0=0$. Then \eqref{eq:princ4} is equivalent to
\begin{align}
&((3\Omega^2-1)(1+x^2)+2(1+\Omega^2)x)y'(x)+\left(\frac{3\Omega^2-1}{x}+1+\Omega^2\right)y(x)=2(3\Omega^2-1)v_1-\frac{2\kappa x}{1-x}\label{eq:diff}
\end{align}
This first-order linear differential equation near $x=0$, can be solved whenever $0<\Omega^2<\frac 13$. Similar considerations apply for $\frac 13\leq\Omega^2\leq1$. One obtains
\begin{align}
&y(x)=\frac{\sqrt{(1-3\Omega^2)(1+x^2)-2(1+\Omega^2)x}}{x}K+\frac{(1-3\Omega^2)v_1}{4\Omega^2(1-\Omega^2)x}(1-3\Omega^2-(1+\Omega^2)x)\nonumber\\
& +\frac{\kappa\sqrt 2}{16\Omega^3x}\arctan\left(\frac{\Omega\sqrt 2(1+x)}{\sqrt{(1-3\Omega^2)(1+x^2)-2(1+\Omega^2)x}}\right) \sqrt{(1-3\Omega^2)(1+x^2)-2(1+\Omega^2)x}\nonumber\\
&+\frac{\kappa(1-3\Omega^2+(\Omega^2-3)x)}{8\Omega^2(1-\Omega^2)x},\label{eq:soldiff}
\end{align}
where $K$ is a constant. As $(v_m)$ is given by the expansion of the solution \eqref{eq:soldiff} near $x=0$, it has to be continous in $x=0$. This fixes the value for $K$. It is given by
\begin{align}
K=-\frac{\sqrt{1-3\Omega^2}}{8\Omega^2(1-\Omega^2)}(2(1-3\Omega^2)v_1+\kappa)-\frac{\kappa\sqrt 2}{16\Omega^3}\arctan\left(\frac{\Omega\sqrt 2}{\sqrt{1-3\Omega^2}}\right).
\end{align}
For $0<\Omega^2<\frac 13$, it is possible to write down the general solution for the $(v_m)$'s. Here we will assume that $\kappa=0$ for the sake of simplicity. From the relation
\begin{align}
\sqrt{1-2\alpha x+x^2}=\sum_{n=0}^\infty \Big(\sum_{k=0}^{\infty}\frac{\Gamma(\frac 32)}{k! \Gamma(n+1-2k)\Gamma(\frac 32-n+k)}(-2\alpha)^{n-2k}\Big)x^n,
\end{align}
we have
\begin{align}
\forall n\geq2\quad v_n=-\frac{(1-3\Omega^2)^2v_1}{4\Omega^2(1-\Omega^2)} \sum_{k=0}^{\infty}\frac{(-2)^{n+1-2k}\Gamma(\frac 32)}{k!\Gamma(n+2-2k)\Gamma(\frac 12-n+k)} \left(\frac{1+\Omega^2}{1-3\Omega^2}\right)^{n+1-2k}.
\end{align}
Using now the definition for the hypergeometric function
\begin{align}
{}_2F_1(a,b;c;z)=\sum_{k=0}^\infty\frac{\Gamma(a+k)\Gamma(b+k)\Gamma(c)}{\Gamma(a)\Gamma(b)\Gamma(c+k)}\frac{z^k}{k!},
\end{align}
we can conclude that
\begin{align}
\forall m\geq0\quad v_{m+1}=&\frac{(1+\Omega^2)^2v_1}{4\sqrt{\pi}\Omega^2(1-\Omega^2)}\frac{\Gamma(3/2)\Gamma(m+3/2)}{\Gamma(m/2+3/2)\Gamma(m/2+2)} \Big(\frac{1+\Omega^2}{1-3\Omega^2}\Big)^m\nonumber\\
& {}_2F_1\left(-\frac m2-\frac 12,-\frac m2-1;-m-\frac 12; \frac{(1-3\Omega^2)^2}{(1+\Omega^2)^2}\right).\label{eq:bigseq}
\end{align}
Notice that for $\frac{(1-3\Omega^2)^2}{(1+\Omega^2)^2}$ small enough, this expression is positive if we choose $v_1\geq0$. Furthermore, ${}_2F_1\left(-\frac m2-\frac 12,-\frac m2-1;-m-\frac 12; z\right)$ is a polynom of degree $\lfloor \frac{m+2}{2}\rfloor$ in $z$, so that:
\begin{align}
v_2&=\frac{1+\Omega^2}{1-3\Omega^2}v_1\nonumber\\
v_3&=\frac{(1+4\Omega^2-4\Omega^4)}{(1-3\Omega^2)^2}v_1\nonumber\\
v_4&=\frac{(1+\Omega^2)(1+8\Omega^2-5\Omega^4)}{(1-3\Omega^2)^3}v_1\nonumber\\
v_5&=\frac{(1+16\Omega^2+26\Omega^4-24\Omega^6-3\Omega^8)}{(1-3\Omega^2)^4}v_1\nonumber\\
v_6&=\frac{(1+\Omega^2)(1+24\Omega^2+66\Omega^4-96\Omega^6+21\Omega^8)}{(1-3\Omega^2)^5}v_1,...
\end{align}
Notice also that using \eqref{eq:soldiff}, the commutative limit for $D=4$ can be obtained in a way similar to what has been done for the 2-dimensional case. The solution $(v_m)$ is indeed continous in $\Omega=0$ for a well-chosen coefficient $v_1=\frac{1}{\theta}+O(\Omega)$ and with $\kappa=O(\Omega)$. It can be realized that the sequence given by \eqref{eq:bigseq} is divergent since it behaves like an exponential so that it does not belong to the Moyal algebra. However, as in the two-dimensionnal case, it is possible to calculate in \eqref{eq:soldiff} the contribution for $\kappa\neq0$ and to set to zero the coefficient of the divergent part for the solution $(v_m)$, so that the resulting vacuum will again belong to the Moyal algebra with suitable asymptotic behaviour.

In the general case, for $\Omega^2\in[0,1]$ and $\kappa\neq0$, we can express the general solution in the configuration space.
\begin{align}
Z_1(x)=-i\sum_{m_1,m_2=0}^\infty \sqrt{m_1+1}a_{m_1,m_2}f_{m_1,m_1+1}(x_1,x_2)f_{m_2,m_2}(x_3,x_4),
\end{align}
where $a_{m_1,m_2}=e^{i\xi_m}\sqrt{v_{m+1}}$, $m=m_1+m_2$ and $\xi_m\in\mathbb R$ is an arbitrary phase. Using the polar coordinates \eqref{eq:polcoord4d} and the expression \eqref{eq:laguerre}, one has
\begin{align}
Z_1(x)=-4i\sqrt{\frac{2}{\theta}}r_1e^{i\varphi_1}e^{-\frac{r^2}{\theta}}\sum_{m=0}^\infty\sum_{m_1=0}^me^{i\xi_m}(-1)^m\sqrt{v_{m+1}}L_{m_1}^1(\frac{2r_1^2}{\theta})L_{m-m_1}^0(\frac{2r_2^2}{\theta}).
\end{align}
Since the identity
\begin{align}
\sum_{k=0}^mL_k^\alpha(x)L_{m-k}^\beta(y)=L_m^{\alpha+\beta+1}(x+y)
\end{align}
is verified and with \eqref{eq:laguerreint}, we find
\begin{align}
Z_1(x)=4\sqrt{\frac{\theta}{2}}\frac{e^{\frac z2}}{z}\int_0^\infty &dt\ e^{-t}J_2(2\sqrt{tz})\sum_{m=0}^\infty \frac{(-1)^m}{m!}\sqrt{v_{m+1}}t^{m+1}((\widetilde x_1+i\widetilde x_2)\cos(\xi_m)\nonumber\\
&+\frac{2}{\theta}(x_1+ix_2)\sin(\xi_m)),\nonumber\\
Z_2(x)=4\sqrt{\frac{\theta}{2}}\frac{e^{\frac z2}}{z}\int_0^\infty &dt\ e^{-t}J_2(2\sqrt{tz})\sum_{m=0}^\infty \frac{(-1)^m}{m!}\sqrt{v_{m+1}}t^{m+1}((\widetilde x_3+i\widetilde x_4)\cos(\xi_m)\nonumber\\
&+\frac{2}{\theta}(x_3+ix_4)\sin(\xi_m)),
\end{align}
where $z=\frac{2x^2}{\theta}$, $v_{m+1}$ is determined above, and $Z_2(x)$ is computed in the same way. The vacuum of the covariant coordinates can therefore be written
\begin{align}
\mathcal A_\mu(x)=2\sqrt{2\theta}\frac{e^{\frac z2}}{z}\int_0^\infty dt\ e^{-t}J_2(2\sqrt{tz})\sum_{m=0}^\infty \frac{(-1)^m}{m!}\sqrt{v_{m+1}}t^{m+1}(\widetilde x_\mu\cos(\xi_m)+\frac{2}{\theta}x_\mu\sin(\xi_m)).\label{eq:vacx4}
\end{align}

\section{Discussion.}

Recent attempts to extend the harmonic solution proposed in \cite{Grosse:2004yu} to the case of gauge theories defined on Moyal spaces have singled out a class of gauge theory models generically described by the action given in \eqref{eq:decadix1} for which the gauge potential has a non vanishing expectation value signaling therefore a non trivial vacuum. In this paper, we have performed a detailled study of the corresponding vacuum states, focussing on those configurations that are invariant under both rotation and symplectic isomorphisms, i.e invariant under $G_D=SO(D)\cap Sp(D)$ which is a symmetry group for the action, as discussed in the section 3. Recall that the explicit determination of these vacua is a necessary step to be reached before the study of its renormalisability can be undertaken since a reliable pertubative analysis in the present situation can only be defined after the action is expanded around the non trivial vacuum. The use of the matrix base for both $D=2$ and $D=4$ dimensions proved very convenient when solving the relevant equations of motion in order to obtain rather tractable expressions, written first in the matrix base and turned back to the position space when necessary. Notice that the technical machinery we set-up in the sections 4 and 5 of this paper provides, as a byproduct, a rather simple algorithm to solve equation of motion that involve Moyal product together with (star)-polynomial interactions.

As the main result of this paper, we have found that the vacuum configurations in the respectively $D=2$ and $D=4$-dimensional position space are generically given
\begin{align}
\mathcal A_\mu^{2D}(x)&=2\sqrt{\theta}\frac{e^{\frac z2}}{\sqrt z}\int_0^\infty dt\ e^{-t}\sqrt tJ_1(2\sqrt{tz}) \sum_{m=0}^\infty \frac{(-1)^m\sqrt{u_{m+1}}}{m!\sqrt{m+1}}t^m\left(\widetilde x_\mu\cos(\xi_m)+\frac{2}{\theta}x_\mu\sin(\xi_m)\right),\\
\mathcal A_\mu^{4D}(x)&=2\sqrt{2\theta}\frac{e^{\frac z2}}{z}\int_0^\infty dt\ e^{-t}tJ_2(2\sqrt{tz})\sum_{m=0}^\infty \frac{(-1)^m}{m!}\sqrt{v_{m+1}}t^{m}\left(\widetilde x_\mu\cos(\xi_m)+\frac{2}{\theta}x_\mu\sin(\xi_m)\right),
\end{align}
where $\xi_m$, $(u_m)$ and $(v_m)$ have been defined in subsections 5.1 and 5.3 and $z=\frac{2x^2}{\theta}$. Note that these solutions do not correspond to the whole set of $G_D$-invariant solutions for the equation of motion. Indeed, we have made in section 4 the assumption $(\mathfrak H)$ that the coefficients $a_m$ and $a_{m_1,m_2}$ are non zero. Let us now discuss this assumption. In fact, it is temptating to conjecture that requiring hypothesis $(\mathfrak H)$ permits one to select only the minima of the action among all the solutions of the equation of motion. This is somewhat supported by the scalar case studied in \cite{deGoursac:2007uv} for which the equation of motion (again a Moyal cubic equation) bears some similarity with the one considered in this paper. In that scalar case, it has been shown that the minima are obtained for a maximal use of assumption $(\mathfrak H)$. Moreover, this is also verified for the two special  cases $\Omega=0$ and $\Omega^2=\frac 13$. Indeed, the solution for $\Omega=0$ corresponds to the usual vacuum $A_\mu=0$, which is of course a minimum of the action, while the case $\Omega^2=\frac 13$ in two dimensions has been treated in the subsection 5.2.  This unfortunately is much more difficult to verify when $\Omega$ is arbitrary because the operator involved in the quadratic part of the action expanded around the vacuum depends then on four indices so that diagonalisation is very difficult (see subsection 5.2). Inclusion of ghost terms into the action stemming from some further gauge fixing might well improve this situation as it can be realized by inspection of the relevant expressions in the matrix base. This would remain to be investigated. In any cases, if the above conjecture was not verified, it is easy to obtain the coefficients $a_m$ and $a_{m_1,m_2}$ of all the solutions of the equation of motion in the matrix base from a rather straighforward adaptation of the results of the section 5. Notice however that this produces a huge number of possible solutions. This will not be considered in the present paper.

Let us finally focus on another special case: $\Omega=1$ and $\kappa<0$. The equation of motion \eqref{eq:motion} simplifies into
\begin{align}
2\mathcal{A}_\mu\star\mathcal{A}_\nu\star\mathcal{A}_\nu + 2\mathcal{A}_\nu\star\mathcal{A}_\nu\star\mathcal{A}_\mu + 2\kappa\mathcal{A}_\mu=0,\label{eq:mot1}
\end{align}
which can be reexpressed in term of the gauge invariant condensate $C(x)=(\mathcal A_\mu\star\mathcal A_\mu)(x)$ as
\begin{align}
\{\mathcal A_\mu,2C+\kappa\}_\star=0.\label{eq:mot2}
\end{align}
It is obvious that the constant condensate
\begin{align}
C(x)=-\frac{\kappa}{2}\label{condensate}
\end{align}
satisfies this equation of motion. The solutions found in the section 5 by requiring the assumption $(\mathfrak H)$ are
\begin{itemize}
\item $u_0=0$, $u_{2k}=-\frac{\kappa}{2}$ and $u_{2k+1}=0$, in two dimensions,
\item $v_0=0$, $v_{2k}=v_{2k+1}=-\frac{\kappa}{8k+4}$, in four dimensions,
\end{itemize}
and all these solutions are of constant condensate type \eqref{condensate}.

\appendix
\section{Appendix}

In this appendix, we will prove recurrently that the coefficients $a_{m_1,m_2}$ depend only on $m=m_1+m_2$ for all $m_1,m_2\in\mathbb N$. We have assumed that $a_{m_1,m_2}\neq0$. Define $b_0=a_{0,0}$. Now suppose that for a certain $m\in\mathbb N$, $\forall k_1,k_2\in\mathbb N$ so that $k_1+k_2\leq m$, $a_{k_1,k_2}$ depends only on $k_1+k_2$ and we write $a_{k_1,k_2}=b_{k_1+k_2}$. Set also $m_1,m_2$ so that $m_1+m_2=m$. Let us prove that $a_{m_1+1,m_2}=a_{m_1,m_2+1}$.

As $a_{m_1,m_2}=a_{m_1+1,m_2-1}=b_m$ and $a_{m_1-1,m_2}=a_{m_1,m_2-1}=b_{m-1}$, equations \eqref{eq:fondam5} can then be reexpressed as
\begin{subequations}
\label{eq:append1}
\begin{align}
&\Big((3\Omega^2-1)m|b_{m-1}|^2+(1+\Omega^2)(2m+3)|b_m|^2+2\kappa+((3\Omega^2-1)(m_1+2)\nonumber\\
&+(1+\Omega^2)(m_2+1))|a_{m_1+1,m_2}|^2 -2(1-\Omega^2)(m_2+1)\overline a_{m_1+1,m_2}a_{m_1,m_2+1} \Big)b_m=0,\label{eq:append1a}\\
&\Big((3\Omega^2-1)m|b_{m-1}|^2+(1+\Omega^2)(2m+3)|b_m|^2+2\kappa+((3\Omega^2-1)(m_2+2)\nonumber\\
&+(1+\Omega^2)(m_1+1))|a_{m_1,m_2+1}|^2 -2(1-\Omega^2)(m_1+1)\overline a_{m_1+1,m_2}a_{m_1,m_2+1} \Big)\overline b_m=0.\label{eq:append1b}
\end{align}
\end{subequations}
If we do the transformation $m_1\to m_1+1$, $m_2\to m_2-1$ in \eqref{eq:append1b} and $m_1\to m_1-1$, $m_2\to m_2+1$ in \eqref{eq:append1b}, and simplify by $b_m\neq0$, we obtain
\begin{subequations}
\label{eq:append2}
\begin{align}
&(3\Omega^2-1)m|b_{m-1}|^2+(1+\Omega^2)(2m+3)|b_m|^2+2\kappa+((3\Omega^2-1)(m_2+1)\nonumber\\
&+(1+\Omega^2)(m_1+2))|a_{m_1+1,m_2}|^2 -2(1-\Omega^2)(m_1+2)\overline a_{m_1+1,m_2}a_{m_1,m_2+1} =0,\label{eq:append2a}\\
&(3\Omega^2-1)m|b_{m-1}|^2+(1+\Omega^2)(2m+3)|b_m|^2+2\kappa+((3\Omega^2-1)(m_1+1)\nonumber\\
&+(1+\Omega^2)(m_2+2))|a_{m_1,m_2+1}|^2 -2(1-\Omega^2)(m_2+2)\overline a_{m_1+1,m_2}a_{m_1,m_2+1} =0.\label{eq:append2b}
\end{align}
\end{subequations}
Then, simplification by $b_m$ in \eqref{eq:append1a} and addition by \eqref{eq:append2a} give rise to the following equation
\begin{align}
&2(3\Omega^2-1)m|b_{m-1}|^2+2(1+\Omega^2)(2m+3)|b_m|^2+4\kappa+4\Omega^2(m+3)|a_{m_1+1,m_2}|^2\nonumber\\ &-2(1-\Omega^2)(m+3)\overline a_{m_1+1,m_2}a_{m_1,m_2+1} =0.\label{eq:append3}
\end{align}
In the same way, with \eqref{eq:append1b} and \eqref{eq:append2b}, we obtain
\begin{align}
&2(3\Omega^2-1)m|b_{m-1}|^2+2(1+\Omega^2)(2m+3)|b_m|^2+4\kappa+4\Omega^2(m+3)|a_{m_1,m_2+1}|^2\nonumber\\ &-2(1-\Omega^2)(m+3)\overline a_{m_1+1,m_2}a_{m_1,m_2+1} =0.\label{eq:append4}
\end{align}
The comparison of \eqref{eq:append3} and \eqref{eq:append4} gives
\begin{align}
|a_{m_1+1,m_2}|^2=|a_{m_1,m_2+1}|^2.\label{eq:square}
\end{align}
By substracting \eqref{eq:append1a} by \eqref{eq:append2b} and using \eqref{eq:square}, we find
\begin{align}
-2(1-\Omega^2)|a_{m_1+1,m_2}|^2+2(1-\Omega^2)\overline a_{m_1+1,m_2}a_{m_1,m_2+1}=0,
\end{align}
and this the aim of the proof:
\begin{align}
a_{m_1+1,m_2}=a_{m_1,m_2+1}.
\end{align}

\vskip 1 true cm 

\noindent
{\bf{Acknowledgments}}: We are grateful to H. Grosse and T. Masson for interesting discussions at various stages of this work. One of us (JCW) gratefully acknowledges partial support from the Austrian Federal Ministry of Science and Research, the High Energy Physics Institute of the Austrian Academy of Sciences and the Erwin Schr\"odinger International Institute of Mathematical Physics   \par

\end{document}